\documentclass[pra,letterpaper,superscriptaddress, amsfonts, 11pt]{revtex4}

\def\be{\begin{equation}}
\def\ee{\end{equation}}
\def\bea{\begin{eqnarray}}
\def\eea{\end{eqnarray}}
\def\bi{\begin{itemize}}
\def\ei{\end{itemize}}
\def\bn{\begin{enumerate}}
\def\en{\end{enumerate}}
\def\ss{\subsection}
\def\sss{\subsubsection}

\usepackage{amssymb}
\usepackage{graphicx}
\usepackage{verbatim}
\usepackage{amsthm}
\usepackage{amsmath}
\usepackage{amsfonts}
\usepackage{url}
\usepackage{amssymb}

\def\T{\mathbf{T}}
\def\bell{\mathbf{l}}
\def\btt{\mathbf{t}}

\def\l{\lambda}
\def\s{\Sigma_{\T}}

\def\S{\Sigma}

\def\t{\theta}
\def\bt{\mathbf{t}}

\newtheorem{Proposition}{Proposition}[section]

\newtheorem{Assumption}{Assumption}[section]

\begin{document}

\bibliographystyle{apsrev}

\title{Evolutionary Inference for Function-valued Traits: Gaussian Process Regression on Phylogenies}
\author{Nick S. Jones}
\affiliation{
Department of Mathematics, Imperial College London, SW7 2AZ}
\author{John Moriarty}
\affiliation{School of Mathematics, University of Manchester, Oxford Road, Manchester M13 9PL, UK}

\begin{abstract}

Biological data objects often have both of the following features: (i) they are functions rather than single numbers or vectors, and (ii) they are correlated due to phylogenetic relationships. In this paper we give a flexible statistical model for such data, by combining assumptions from phylogenetics with Gaussian processes. We describe its use as a nonparametric Bayesian prior distribution, both for prediction (placing posterior distributions on ancestral functions) and model selection (comparing rates of evolution across a phylogeny, or identifying the most likely phylogenies consistent with the observed data). Our work is integrative, extending the popular phylogenetic Brownian Motion and Ornstein-Uhlenbeck models to functional data and Bayesian inference, and extending Gaussian Process regression to phylogenies. We provide a brief illustration of the application of our method.

\end{abstract}

\maketitle
\section{Introduction}

In this paper we consider statistical inference for function-valued data which are correlated due to phylogenetic relationships. A schematic example is given in Figure 1A: in this case, given functional data observed at the tips of a phylogeny, the task is to perform inference on the (unobserved) functional data at the root of the phylogeny. Alternatively, if the phylogeny is uncertain we may wish to perform phylogenetic inference, or our interest may be inferring the dynamics of the evolutionary process which produced the data. The term `function-valued' is meant in the sense of \cite{Ramsay82}, where a datum is a continuous function $f(x)$ of a variable $x$, such as time or temperature: an examples are therefore curves for ambient temperature versus growth rate for caterpillars, a heart rhythm time series~\cite{ramsay2005functional}, or a spectrogram of audio data. Our approach is to combine the theory of Gaussian processes with assumptions from phylogenetics, to obtain a flexible nonparametric model for such data. Since this model effectively specifies the evolutionary dynamics of the data through the phylogeny, we note that (i) our approach generalises the Brownian Motion and Ornstein-Uhlenbeck models of continuous-time character evolution from quantitative genetics \cite{Lande76}, and (ii) the phylogenetic tree will play the role of evolutionary time: to avoid confusion, we therefore refer to the indexing variable $x$ above as `space'. The model may be used as a prior for Bayesian inference, which opens up functional and phylogenetically aware approaches to both prediction and model selection.

Because of their generality, flexibility and mathematical simplicity, there has been substantial recent interest in the use of Gaussian process priors for Bayesian nonparametric regression \cite{rasmussen2006gaussian,Stein99,patil1993multivariate}.
This paper may be viewed as an extension of Gaussian process regression to take account of functional data and a tree topology. 
Our work relates to the field of spatial statistics, in that it involves multidimensional index sets with both distance and topology: however, it is the tree topology and the conditional independence of siblings given their common ancestors that makes our approach particularly suited to the study of phylogenetically correlated data.

Functional representations of data have been in use for at least 30 years \cite{Ramsay82, King}. Their use in (non-phylogenetic) evolutionary studies was proposed for quantitative genetics in \cite{Kirkpatrick89}. While the wider debate concerning classical versus functional data types is outside the scope of the current report, in \cite{Ramsay82} it is argued that statistical challenges for classical data have corresponding dual statistical challenges for functional data, so we may ask: in the application of our models to statistical inference, which classical approaches are dual to our functional approach? Classical data types investigated in the phylogenetic context include sequences of discrete symbols (e.g. encoding the presence or absence of certain characters \cite{Wiens01}, or alternatively in models for the evolution of genetic sequences), the spatial locations of a number of fixed landmark points in geometric morphometrics \cite{Catalano10}, and multivariate vectors of continuous characters or summary statistics \cite{Fel1985}. Within the evolutionary study of multivariate vectors, the relative effects of phylogenetic and (species-)specific variation have been studied \cite{Cheverud85}, which relates closely to our equation \eqref{fullcov} below, and the challenges of prediction and model selection have been addressed and explored statistically \cite{Hansen96, Butler04,Mar1997}. We address model selection in the supplement to this report and more fully  in a companion paper \cite{pantelis}, which also contains a discussion of statistical issues and performance. Beyond this duality, however, our model is suitable for use as a prior in Bayesian analysis, and in the remainder of this paper we derive our models and provide details for their use in Bayesian regression.

\section{Phylogenetic Gaussian processes}
\label{pgps}

\subsection{Gaussian Processes and Regression}\label{suba}
A Gaussian process is a collection of random variables, any finite number of which have a joint Gaussian distribution. Examples are the Wiener (Brownian Motion) and Ornstein-Uhlenbeck processes which have received considerable recent attention in the study of evolution (see, for example, \cite{Fel1985, Butler04, Hansen96, Diniz01}). It is characterised by its first two (cross-)moments: unless otherwise specified, all Gaussian processes in this paper are assumed to have mean zero (which is a common assumption for Gaussian process regression analyses, see \cite{rasmussen2006gaussian}), so the only choice when specifying a Gaussian process is thus how the sample values covary which is encoded by a covariance function $\sigma$, typically depending on a vector $\theta$ of parameters.

Suppose that a Gaussian process $f$ is observed at a vector of co-ordinates $L$. Then the resulting vector of sample values $f(L)$ has
a multivariate Gaussian distribution of dimension equal to $|L|$, the number of points of measurement: $f(L)\sim\mathcal{N}(0,\sigma(L,L,\theta))$. Here $\mathcal N$ represents the Gaussian distribution, its two arguments being the mean vector and the covariance matrix $\sigma(L,L,\theta)$, which is the matrix of the covariances between all pairs $(\ell_i,\ell_j)$ of observation co-ordinates in $L$ (where $\ell_i\in L$) so that
\begin{equation}
[\sigma(L,L,\theta)]_{ij}=\sigma(l_i,l_j,\theta) = E[f(l_i)f(l_j)].
\end{equation}
In subsection \ref{subd} below we derive the structure of some particular covariance functions, including the phylogenetic Ornstein-Uhlenbeck process.

The log-likelihood of the sample $f(L)$ is then \cite{rasmussen2006gaussian}
\begin{equation}\label{log1}
\log p(f(L)|\theta) = -\frac{1}{2}f(L)^T \sigma(L,L,\theta)f(L) - \frac{1}{2} \log(det(\sigma(L,L,\theta)))- \frac{|L|}{2} \log 2\pi.
\end{equation}

We might be interested in making inferences about the unobserved values of our random function $f$ at a vector $M$ of co-ordinates, given samples at the co-ordinates $L$. The posterior distribution of the vector $f(M)$ given $f(L)$ is also Gaussian and of the form \cite{rasmussen2006gaussian}:
\begin{equation}\label{dangermouse}
f(M)|f(L)  \sim \mathcal N (A,B)
\end{equation}
where
\begin{eqnarray}
A& =& \sigma(M,L,\theta)\sigma(L,L,\theta)^{-1}f(L), \label{boselecta} \\
B &=& \sigma(M,M,\theta) - \sigma(M,L,\theta)\sigma(L,L,\theta)^{-1}\sigma(M,L,\theta)^{T} \label{chas and dave}
\end{eqnarray}
and $\sigma(M,L,\theta)$ denotes the $|M| \times |L|$
matrix of the covariance function $\sigma$ evaluated at all pairs $m_i \in M, l_j \in L$. From Eq. (\ref{boselecta}) the posterior mean vector $A$ consists of linear combinations of the observations while the posterior covariance matrix $B$, given by \eqref{chas and dave}, is independent of the observations. Gaussian process regression is nonparametric in the sense that no assumption is made about the structure of the model: the more data gathered, the longer the vector $f(L)$, and the more intricate the posterior model for $f(M)$. 

We are able to combine evolutionary dynamics with functional data because the index variable $\ell$ introduced above can be a point in a space of
arbitrary dimension. Therefore, if we wish to model the time evolution of a functional trait $f(x)$ with indexing variable $x$, we could consider each point of observation $\ell$ as corresponding to a point $(x,t)$ in both space and evolutionary time. Then $f(L)$ would represent the values of a random space-time surface at various space-time co-ordinates, $L$ (analogously, the values $f(L)$ are like a set of altimeter recordings recorded at different locations ($L$) on a map). A cross-section through this space-time surface at a fixed value of time $t$ yields a random curve which can be viewed as a single function-valued trait at the fixed time $t$ in its evolution. In the next section we will extend this view to processes on phylogenies.

\label{aaa}

\subsection{Phylogenetic covariance function}\label{subb}

Our aim in this subsection is to build a Gaussian process model for the evolution of a function-valued trait along a phylogenetic tree $\T$ by allowing $\T$ to play the role of evolutionary time, rather than the linear time variable $t$ used above. Suppose, therefore, that each observation $\ell$  corresponds to a point $(x,\btt)$ in the {\em space-phylogeny} $S \times \T$: that is, $x \in S$ is the value  under consideration of the spatial (indexing) variable, and $\btt \in \T$ is the point under consideration on the phylogeny ($\btt $ is not just a time co-ordinate but also indicates a branch of the phylogeny). We will do this by constructing a covariance function $\s(\bell_i,\bell_j)$ when the $\bell_{i}$, $\bell_{j}$ are points in $S \times \T$, calling it the {\em phylogenetic covariance function}. In order to obtain a unique phylogenetic covariance function $\s$ we will make two assumptions which are natural in the context of evolution (see, for example, \cite{Fel1985}):
\begin{Assumption} Conditional on their common ancestors in the phylogenetic tree $\T$, any two traits are statistically independent.
\label{leighton}
\end{Assumption}
\begin{Assumption} The statistical relationship between a trait and any of its descendants in $\T$ is independent of the topology of $\T$. \label{diniyar2}
\end{Assumption}
Assumption \ref{diniyar2} means that our statistical model of the evolutionary process is identical along paths through $\T$ from the root to any tip,
and we call this the {\em marginal process} (this assumption could be generalised, for example to model unequal rates of evolution along different branches of $\T$.) As in \cite{Fel1985} and related work, we need the assumption that $\T$ is a chronogram, having both a tree topology and a distance metric. We will call the distance between a point $\btt \in \T$ and the root of $\T$ the `date' of $\btt$, and denote it by the plain typeface symbol $t$.

The marginal process is of course Gaussian, and so it is sufficient to specify its covariance function $\Sigma$ on $S \times T$, where $T$ is the set of all dates in $\T$. In fact, each different marginal covariance function specifies a different phylogenetic covariance function, and so a wide class of phylogenetic covariance functions may be constructed. In turn, these models offer an equally wide choice of priors for Bayesian evolutionary inference with function-valued traits.

In the following Proposition we present our mathematical results in the simplest case when the marginal covariance function is space-time separable: that is, when there exist a space-only covariance function $K(x_{1},x_{2})$ and a time-only covariance function $k(t_{1},t_{2})$ such that
\begin{eqnarray} \label{separa}
\S((x_{1},t_{1}),(x_{2},t_{2}))=K(x_{1},x_{2}) k(t_{1},t_{2}).
\end{eqnarray}

\begin{Proposition}\label{prop1}
\bn
\item If the marginal covariance function $\S$ is space-time separable then the phylogenetic covariance function $\s$ is also space-time separable, i.e.
\begin{equation}
\label{pj1}
\s((x_{1},\btt_{1}),(x_{2},\btt_{2}))=K(x_{1},x_{2}) k_{\T}(\btt_{1},\btt_{2})
\end{equation}
where $k_{\T}(\btt_{1},\btt_{2})$ is the phylogenetic covariance function constructed from the time-dependent component $k$ in \eqref{separa} and $K(x_{1},x_{2})$ is the space-dependent component.
\item \label{whattimeislove} When the time-dependent component $k$ of \eqref{separa} specifies a process that is Markovian in time, we have the simple expression
\begin{equation}\label{goldie2}
k_{\T}(\btt_{1},\btt_{2})=k(t_{1},t_{12})k(t_{12},t_{12})^{-1}k(t_{2},t_{12})
\end{equation}
where $\btt_{12}$ is the most recent common ancestor of $\btt_{1}$ and $\btt_{2}$ (and $t_{12}$ is its depth in $\T$). In particular, we have the following corollaries:
\begin{enumerate}
\item if $k(t_{1},t_{1 2})=\min(t_{1},t_{1 2})$ so that we have a Wiener process in evolutionary time as in \cite{Fel1985}, then $k_{\T}(\btt_{1},\btt_{2})=t_{{12}}$, which is the variance of the evolutionary time component of variation evaluated at the most recent common ancestor $\btt_{{12}}$.
\item if $k$ is isotropic so that $k(t_{1},t_{2})$ is a function of $|t_{1}-t_{2}|$ only, it does not necessarily follow that $k_\T(\btt_{1},\btt_{2})$ is isotropic (meaning a function of the patristic distance between $\btt_{1}$ and $\btt_{2}$ only). In fact, $k_\T$ is only isotropic when $k$ is the Ornstein-Uhlenbeck covariance.
\label{to see}
\end{enumerate}
\item Let $Y$ be a phylogenetic Gaussian process with a space-time separable covariance function $\Sigma_\T$ which factorises as in \eqref{pj1}. If $K$ is a continuous degenerate Mercer kernel then there exist an integer $n$ and deterministic functions $\phi_i: S \to \mathbb{R}$ and univariate Gaussian processes $X_i$, for $i=1\ldots n$, such that the Gaussian process given by
\begin{equation}\label{basisrep1}
f(x,\bt) = \sum_{i=1}^{n}\phi_{i}(x)X_{i}(\bt)
\end{equation}
has the same distribution as $Y$.\label{baz}
\end{enumerate}
\end{Proposition}
Further mathematical detail can be found in the supplement. Part \ref{to see} of Proposition \ref{prop1} helps clarify the relationship between phylogenetic Gaussian process models and studies such as \cite{Gittleman90} based on autocorrelation functions of patristic distance, namely that the two are compatible only when the phylogenetic Ornstein-Uhlenbeck process is assumed. Part \ref{baz} establishes that a convenient expansion using basis functions can be used for a wide range of phylogenetic Gaussian processes with space-time separable covariance functions. This expansion is useful for statistical inference, as it justifies the use of dimension reduction techniques (see companion paper \cite{pantelis}).

\begin{flushleft}
\begin{figure}[!ht]
\includegraphics[width=1.0\textwidth]{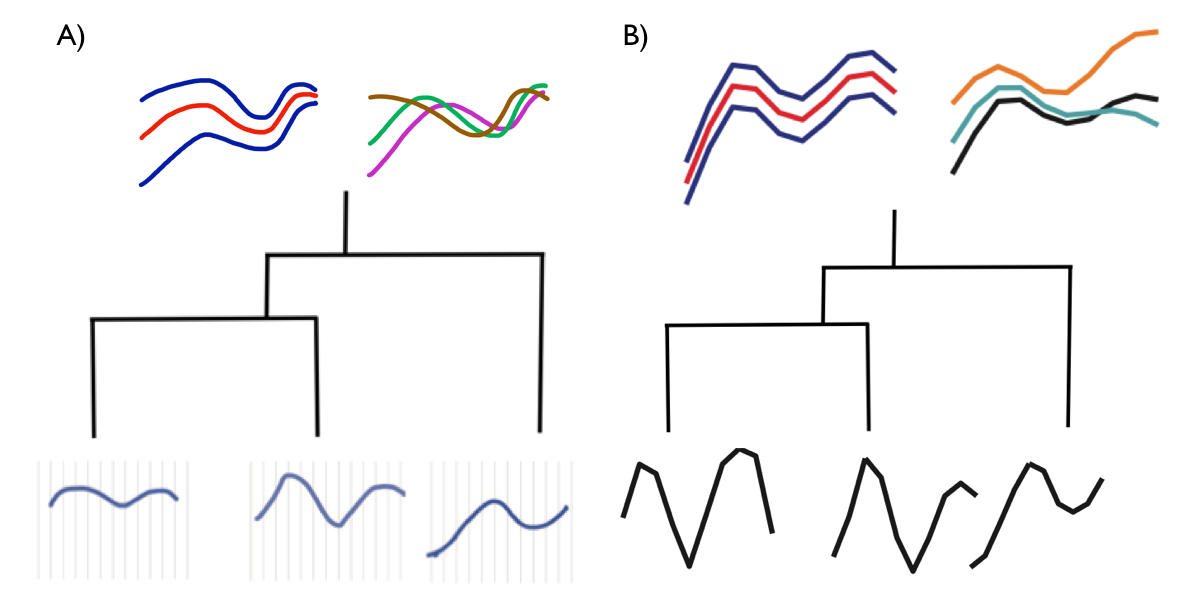}
\caption{Schematic illustration for Bayesian inference on functional data related by a phylogeny, giving a posterior distribution for the function at the root. The red curve at the root indicates the predicted mean surface given only the data from the tips, the blue curves would be one standard deviation uncertainties at each point. The three curves at the top right are notional samples from the Gaussian process at the root. A: spatially inhomogeneous modelling, where the prior is fitted to the observed spatial variation in the sample (see \cite{pantelis} for detail). B: spatially homogeneous modelling, where the prior is spatially homogeneous (see the supplement for detail).}
\label{fig1}
\end{figure}
\end{flushleft}

\ss{Examples}\label{subd}
We now illustrate two ways in which Proposition \ref{prop1} may be used to make priors for Bayesian inference on phylogenetically related functional traits. They differ in their approach to modelling spatial variation, and may be called the spatially inhomogeneous (figure \ref{fig1}A) and spatially homogeneous (\ref{fig1}B) models respectively. In both cases we   assume the simplest possible structure for the marginal covariance function $\S$: that it is space-time separable as in \eqref{separa}. We also assume that, conditional on any given trait, its ancestor and progenitor traits are statistically independent. This corresponds to choosing a temporal component which is Markovian. The only Markovian Gaussian processes are the class of Ornstein-Uhlenbeck processes, and the stationary examples have the covariance function
$
k(t_{1},t_{2})=\exp \left(-|t_{1}-t_{2}|/\theta_2\right)
$
where the hyperparameter $\t_2$ specifies the characteristic length scale for the evolutionary dynamics.
From Proposition \ref{prop1}
we obtain
\begin{equation}
k_\T(t_{1},t_{2})=\exp \left(-d_{\T}(\btt_{1},\btt_{2})/\t_2\right)\label{oupcov}
\end{equation}
where $d_{\T}(\btt_{1},\btt_{2})$ denotes the patristic distance between $\btt_{1}$ and $\btt_{2}$ (note that the phylogenetic Ornstein-Uhlenbeck model {\em is} isotropic, see end of Proposition \ref{prop1}).
\sss{Spatially homogeneous model}\label{unb}
If our prior belief is that variation in the functional trait is homogeneous over all values of $x \in S$ then we should choose the spatial covariance $K$ to be stationary, and if the traits are typically smooth we should choose $K$ to generate smooth random functions. An example is the squared exponential covariance function:
\begin{eqnarray}
K(x_{1},x_{2})&=&\exp\left(-(x_{1}-x_{2})^{2}/2\theta_1^2 \right).\label{secov}
\end{eqnarray}
Here the hyperparameter $\t_1$ fixes the characteristic length scale of the random functions. This choice of $K$ gives
\begin{eqnarray}\label{sepz}
\s((x_{1},\btt_{1}),(x_{2},\btt_{2}))&=&K(x_{1},x_{2}) k_{\T}(\btt_{1},\btt_{2})\\
&=&\exp\left(-(x_{1}-x_{2})^{2}/2\t_1^2-d_{\T}(\btt_{1},\btt_{2})/\t_2\right). \label{gb}
\end{eqnarray}

This simple example may be combined by summation to construct other spatially homogeneous phylogenetic covariance functions, although the separability property is typically then lost. The following phylogenetic covariance function, $\s'$, contains an uncorrelated noise term whose influence is controlled by the choice of the parameter $\sigma^2_n$: \begin{eqnarray}\label{fullcov}
\s'((x_{1},\btt_{1}),(x_{2},\btt_{2}))&=&(1-\sigma^2_n)\exp\left(-(x_{1}-x_{2})^{2}/2\t_1^2-d_{\T}(\btt_{1},\btt_{2})/\t_2\right)+\sigma^2_n\delta_{\btt_{1},\btt_{2}}\delta_{x_1,x_2}
\end{eqnarray}
where $\delta$ is the Kronecker delta. When functional data are sampled at a finite set of space-time points $L$, the phylogenetic covariance function \eqref{gb} may be used as a prior for Bayesian regression as described in subsection \ref{suba}. Figure \ref{fig1}B gives a schematic representation of functional data and the posterior distribution for the root function, when all functions are discretely sampled on a regular lattice. An illustrative example developing Figure \ref{fig1}B is supplied in the supplement.

\sss{Spatially inhomogeneous model}

We may alternatively construct a prior using the representation \eqref{baz}. This involves choosing deterministic spatial basis functions $\phi_1,\ldots,\phi_n:S \to \mathbb{R}$ and univariate phylogenetic Gaussian processes $X_1,\ldots,X_n$. The spatial basis may be specified a priori, or alternatively obtained by functional decomposition of observed data: in the companion paper \cite{pantelis} we present a practical methodology for this empirical Bayesian approach. If we assume that the $X_i$ are independent phylogenetic Ornstein-Uhlenbeck processes on $\T$ with noise, then we have
\begin{equation}
k_\T^i(\btt_1,\btt_2) = (1-(\sigma_n^i)^2)\exp \left(-d_{\T}(\btt_{1},\btt_{2})/\t_2\right) + (\sigma_n^i)^2 \delta_{\btt_1,\btt_2}.
\end{equation}
The full covariance function of our prior distribution is then
\bea
\label{pj}
\s^{\text{inhom}}((x_{1},\bt_{1}),(x_{2},\bt_{2}))
=E\left[f(x_{1},\bt_{1})f(x_{2},\bt_{2})\right]&=&\sum_{i=1}^{n}k_{\T}^i(\bt_{1},\bt_{2})\phi_{i}(x_{1})\phi_{i}(x_{2}).
\eea
Figure \ref{fig1}A gives a schematic representation of functional data and the posterior distribution for the root function, when all functions are constructed from the spatial basis $\phi_1,\ldots,\phi_n$.

\section{Discussion} \label{theone}

In this paper we have exploited the powerful inference architecture provided by Gaussian processes to address phylogenetic questions for function-valued data.
Explicit posterior distributions are available, giving a straightforward approach to the prediction of unobserved function-valued traits, as well as a principled approach to evolutionary model selection. The approach is suitable both for complete (or dense) observations of function-valued traits and for sparsely and even irregularly sampled traits, with missing observations.
\bibliographystyle{vancouver}

\end{document}